*Article*

# The Opacity Project: R-Matrix Calculations for Opacities of High-Energy-Density Astrophysical and Laboratory Plasmas

Anil K. Pradhan [1,2,*] and Sultana N. Nahar [1]

1. Department of Astronomy, The Ohio State University, Columbus, OH 43210, USA; nahar.1@osu.edu
2. Chemical Physics Program, The Ohio State University, Columbus, OH 43210, USA
* Correspondence: pradhan.1@osu.edu; Tel.: +1-614-292-5850

**Abstract:** Accurate determination of opacity is critical for understanding radiation transport in both astrophysical and laboratory plasmas. We employ atomic data from R-Matrix calculations to investigate radiative properties in high-energy-density (HED) plasma sources, focusing on opacity variations under extreme plasma conditions. Specifically, we analyze environments such as the base of the convective zone (BCZ) of the Sun ($2 \times 10^6$ K, $N_e = 10^{23}$/cc), and radiative opacity data collected using the inertial confinement fusion (ICF) devices at the Sandia Z facility ($2.11 \times 10^6$ K, $N_e = 3.16 \times 10^{22}$/cc) and the Lawrence Livermore National Laboratory National Ignition Facility. We calculate Rosseland Mean Opacities (RMO) within a range of temperatures and densities and analyze how they vary under different plasma conditions. A significant factor influencing opacity in these environments is line and resonance broadening due to plasma effects. Both radiative and collisional broadening modify line shapes, impacting the absorption and emission profiles that determine the RMO. In this study, we specifically focus on electron collisional and Stark ion microfield broadening effects, which play a dominant role in HED plasmas. We assume a Lorentzian profile factor to model combined broadening and investigate its impact on spectral line shapes, resonance behavior, and overall opacity values. Our results are relevant to astrophysical models, particularly in the context of the solar opacity problem, and provide insights into discrepancies between theoretical calculations and experimental measurements. In addition, we investigate the equation-of-state (EOS) and its impact on opacities. In particular, we examine the "chemical picture" Mihalas–Hummer–Däppen EOS with respect to level populations of excited levels included in the extensive R-matrix calculations. This study should contribute to improving opacity models of HED sources such as stellar interiors and laboratory plasma experiments.

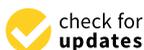

## 1. Introduction

HED plasma environments span a huge variety of astrophysical and laboratory sources. Many recent efforts have focused on radiation transport in many astrophysical objects, and laboratory experiments have focused on specific ranges of temperatures and densities. The fundamental parameter that encompasses and underlies radiation transport is the opacity of the plasma in the source.

Of special interest is the Sun as a benchmark in the laboratory for astrophysical processes. This includes precise determination of absolute abundances X/H of elements heavier than helium, interior plasma opacity, sound speed, and primordial helium abundance. One of the main advances in addressing the solar problem is that helioseismic data yields direct information on these and related parameters and processes ([1] and references therein). Surprisingly, however, these important parameters are brought into question owing to a large discrepancy between solar abundances determined from new spectroscopic observations and elaborate Non-LTE (non-local-thermodynamic-equilibrium) models, which determine elemental abundances of light volatile elements, such as carbon, nitrogen, and oxygen, to be much lower than the previous "standard" solar abundances [2]. It has been suggested that an increase in solar interior opacities could account for lower abundances [3] and problems in helioseismology [1,4]. Therefore, a re-examination of current astrophysical opacities with higher accuracy is necessary.

Following the inception of the Opacity Project in 1983 (OP; [5–11]), large-scale atomic calculations using the powerful, state-of-the-art R-matrix method [9] were envisaged to compute high-accuracy opacities. But despite enormous computational and theoretical effort, the complexity of R-matrix calculations proved to be intractable until the past decade (Nahar and Pradhan 2016, hereafter NP16 [12,13]). The problems encountered earlier have since been overcome to a large extent, and monochromatic and mean R-matrix opacities can now be computed for complex ions of iron prevalent around the solar BCZ [14–18]. About half the opacity at the BCZ is due to only two elements, oxygen and iron. But around BCZ temperatures and densities, oxygen is either fully ionized or H-like O VII, and is therefore not a significant contributor per atom. As such, the overall enhancement in solar interior opacity is likely due to the abundance of iron in multiple ionization states with active L-shell electrons [14].

Concurrently, considerable effort has been been devoted to recreating solar interior plasma conditions in the laboratory, and to measure opacity spectra at different temperatures and densities. In particular, the inertial confinement fusion (ICF) device at the Sandia National Laboratory has reported measurements of iron opacity that are higher than theoretically predicted values [19]. However, the experimental opacities for Cr and Ni do not appear to show such enhancement in the Sandia Z experiments [20]. In addition, recent radiation burn-through measurements of iron opacity at the Lawrence Livermore Laboratory do not indicate a large increase in Fe opacity under BCZ conditions [21]. However, more recent results of experimental works at the Livermore, Sandia, and Los Alamos laboratories may be found in a report related to the Sandia Z machine and the National Ignition Facility at Livermore [22] (see Section 5 on comparison with experimental data). Thus, an accurate recalculation of theoretical opacities is also needed to resolve this apparent discrepancy between experimental data for Fe opacity, as well as for other heavy elements.

More generally, opacities related to the solar problem have a bearing on fundamental issues in high-energy-density (HED) plasma physics. These include the equation-of-state (EOS) of HED plasmas in local thermodynamic equilibrium (LTE). In particular, we examine the Mihalas–Hummer–Däppen EOS used in the OP and variants thereof, together with possible limitations that would affect opacity calculations. Another important issue concerns plasma effects on resonant phenomena in atomic processes, as reflected in photoionization cross-sections that determine bound–free opacity [23].

In this presentation we describe the R-matrix methodology for opacity calculations with illustrative results pertaining to the topics mentioned above.

## 2. R-Matrix Opacity Calculations

The OP adapted the R-matrix method from atomic collision theory, which has proven to be most successful in treating a variety of atomic processes such as (e + ion) scattering, radiative transitions, photoionization, and (e + ion) recombination [9,10,24]. The R-matrix (RM) method, and its relativistic version, the Breit–Pauli R-matrix (BPRM) [24], is based on the coupled channel (CC) approximation, which may be employed to compute total wavefunctions for both bound- and continuum-free (e + ion) states. Therefore, the RM method enables and is equivalent to the extensive investigation of configuration interaction and electron correlation effects. Primary among these is the precise delineation of autoionizing resonances that are ubiquitous in (e + ion) collisions and photoionization.

Under the OP, the RM method was extended to compute the huge amount of atomic data needed for opacity calculations [5,6,8]. However, most of the earlier calculations were carried out with the non-relativistic version in LS coupling. More recently, within the past decade, Systematic computations have been carried out using the BPRM version [12,13], as described in [14–17] and in this paper.

The CC approximation entails the following wavefunction expansion for the (e + ion) system:

$$\Psi_E(e + ion) = A \sum_i \chi_i(ion)\theta_i + \sum_j c_j \Phi_j, \quad (1)$$

where the atomic system is represented as the 'target' or the 'core' ion of N-electrons interacting with the $(N+1)^{th}$ electron. The $(N+1)^{th}$ electron may be bound in the electron–ion system, or in the electron–ion continuum, depending on whether its energy is negative or positive. The total wavefunction, $\Psi_E$, of the (N+1)-electron system in a symmetry, $SL\pi$ or $J\pi$, is an expansion over the eigenfunctions of the target ion, $\chi_i$, in a specific state $S_iL_i(J_i)\pi_i$, coupled with the $(N+1)^{th}$ electron function, $\theta_i$. The sum is over the ground and excited states of the target or the core ion. The $(N+1)^{th}$ electron with kinetic energy $k_i^2$ corresponds to a channel labeled $S_iL_i(J_i)\pi_i k_i^2 \ell_i(SL(J)\pi)$. The $\Phi_j$s are bound channel functions of the (N+1)-electron system that account for short-range correlation. The total wavefunction expansion represents a continuum wavefunction $\Psi_F$ for an electron with positive energies (E > 0), or a bound state $\Psi_B$ with a *negative* total energy (E ≤ 0). The complex resonance structures in photoionization cross-sections result from channel couplings between the continuum channels that are open ($k_i^2 > 0$) and the ones that are closed ($k_i^2 < 0$). In addition to an accurate wavefunction representation of the (e + ion) system, the CC approximation, as employed in the RM method, gives rise to autoionizing resonances owing to quantum interference or coupling among the bound and continuum channel components. Resonances occur at electron energies $k_i^2$ corresponding to autoionizing states belonging to Rydberg series $S_iL_i\pi_i\nu\ell$, where $\nu$ is the effective quantum number converging onto the target threshold $S_iL_i$.

The computer codes related to relativistic and non-relativistic RM calculations are described in [5,6,9,10,14,24]. The first step is the construction of an accurate target eigenfunction expansion using the atomic structure code Superstructure, CIV3, or Grasp [10,25]. The second step is the computation of algebraic coefficients for the (e + ion) system, followed by the third step of diagonalization of the (e + ion) Hamiltonian that determines the R-matrix basis functions. Subsequent steps utilize these basis functions to compute "asymptotic" quantities such as bound and continuum (e + ion) wavefunctions, collision cross-sections, photoionization cross-sections, energy levels, and transition probabilities [10].

In contrast to the CC approximation and the BPRM method, the distorted wave (DW) approximation employed in other opacity models neglects channel coupling, and therefore, resonances are not included in an ab initio manner (although they may be considered perturbatively). The practical consequence of resonances in atomic cross-sections is that autoionization broadening is included precisely to obtain *intrinsic* widths in bound–free absorption and opacity. A theoretical and computational treatment has now been developed to consider *extrinsic* plasma effects in RMOP-III [16], as discussed later.

Prior to the recent RMOP work that is the focus of this paper, existing OP calculations mainly employed atomic-structure calculations followed by DW approximation as in other opacity models, although the OP also included a limited number of non-relativistic R-matrix calculations for outer open-shell configurations. The OP atomic data and opacities have been computed and archived in the online databases TOPbase, OPserver, and NORAD-Atomic-Data [26–29].

In an earlier study, Delahaye et al. (2021; hereafter D21 [30]) presented Dirac R-matrix calculations for Fe XVII photoionization cross-sections up to n = 5, 6 configurations for the core ion Fe XVIII. The D21 results found convergence with respect to the excited *n*-complexes, demonstrated in an earlier study with respect to *n* = 2, 3, 4 core levels [12,13]. However, D21 did not directly compare their RM cross-sections with NP16 cross-sections, and did not consider temperature–density-dependent plasma broadening effects on resonances in RM cross-sections, as described in this paper. Most likely, owing to neglect of plasma broadening effects on autoionizing resonances in photoionization cross-sections, D21 also obtained lower RMOs from their RM than the present RMOP and other calculations.

Illustrative results from RMOP-IV are shown in Figure 1, comparing CC relativistic Breit–Pauli R-matrix (BPRM) calculations and relativistic DW (RDW) photoionization cross-sections. Whereas the agreement between the BPRM and RDW background cross-sections is good, the situation is uncertain within the range of resonances that cannot be calculated in RDW approximation in ab initio manner (but may be included perturbatively). The CC BPRM calculations entail large-scale eigenfunction expansions for Fe XVII with 218 target (core) levels of Fe XVIII in the (e + Fe XVIII) system, and 276 core levels of Fe XIX for Fe XVIII. Resonances in BPRM results often span very wide autoionization widths. They correspond to energies of strong dipole transitions in the core ion and are referred to as photoexcitation-of-core (PEC) or Seaton resonances, which may dominate the bound–free opacity [8,10].

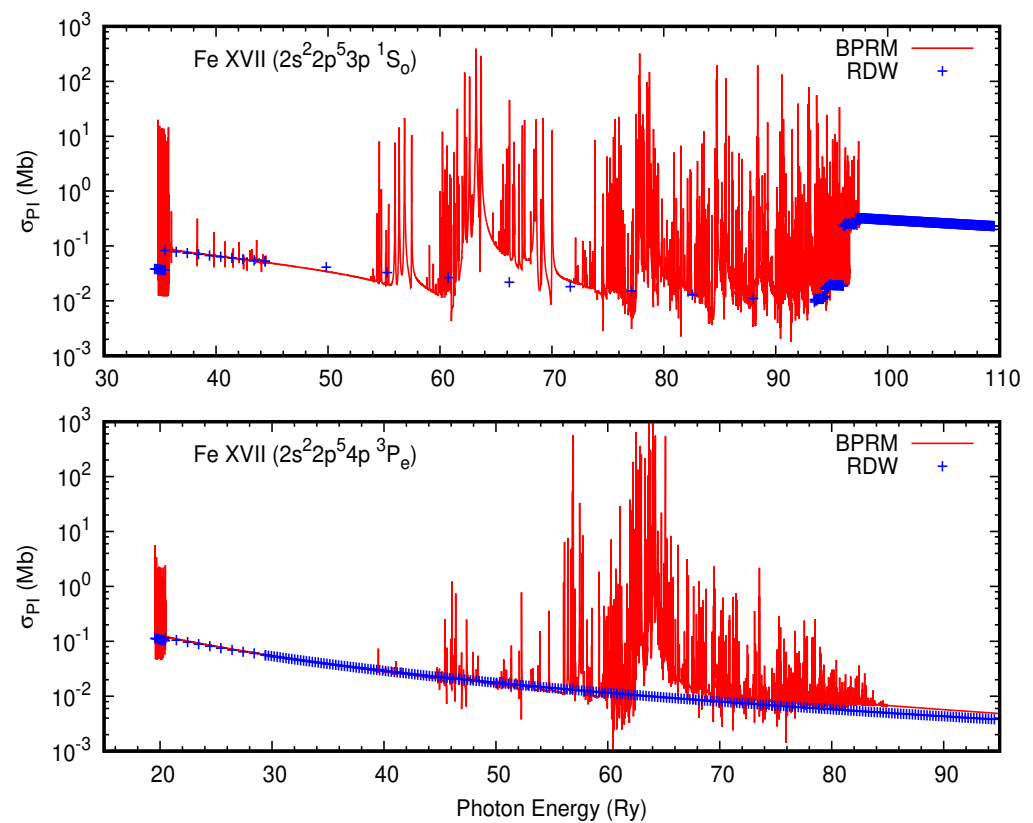

**Figure 1.** Comparison between Breit–Pauli R-matrix (BPRM) and Relativistic Distorted Wave (RDW) photoionization cross-sections for two excited levels of Fe XVII: Top—$2s^22p^53p\ ^1S^e$; Bottom—

$2s^22p^54p\ ^3P^e$. The RDW data (blue) reproduce the background cross-sections well, but not the extensive resonance structures in the BPRM cross-sections with large autoionization widths (red), particularly the broad and asymmetric photoexcitation-of-core (PEC) resonances that dominate bound–free continuum opacity. Although the RDW and BPRM background cross-sections match, they often need scaling up or down in order to coincide.

## 3. Plasma Broadening

As evident from the previous section, one of the main differences between RMOP and DW opacities is due to electron correlation effects that particularly manifest in photoionization cross-sections. However, in addition, plasma effects on the intrinsically broadened autoionization resonances in Figure 1 must be taken into account. Whereas line-broadening treatments are well-known and implemented in opacity models, including earlier OP work [8], the extrinsic plasma broadening treatment of autoionizing resonances in CC R-matrix calculations for atomic processes has only recently been developed [16]. The new method for plasma broadening of autoionizing resonances incorporates the primary mechanisms due to (I) the thermal Doppler effect, (II) Stark ion microfields, (III) electron collisions, and (IV) free–free transitions. Unbroadened R-matrix cross-sections with energy-delineated resonance structures are convolved over a Lorentzian functional at each temperature–density combination [16].

The Lorentizan convolution of a photoionization cross-section of a level is expressed as

$$\sigma_i(\omega) = \int \tilde{\sigma}(\omega')\phi(\omega',\omega)d\omega', \qquad (2)$$

where the unbroadened cross-sections with AI resonance structures are denoted as $\tilde{\sigma}$, and plasma-broadened cross-sections as $\sigma$. The photon energy $\omega$ is in Rydberg atomic units, and $\phi(\omega',\omega)$ is the normalized profile factor in terms of the total width $\Gamma$ including all broadening processes:

$$\phi(\omega',\omega) = \frac{\Gamma(\omega)/\pi}{x^2 + \Gamma^2}, \qquad (3)$$

where $\omega - \omega' \equiv x$. The difference with line broadening is that resonances correspond to quantum mechanical interference between continua, defined by excited core ion levels and corresponding (e + ion) channels. The RM method yields resonance profiles that are generally asymmetric, unlike line profiles, which are usually symmetric. With $N$ core ion levels, we have

$$\sigma(\omega) = \sum_i^N \left[ \int \tilde{\sigma}(\omega') \left[ \frac{\Gamma_i(\omega)/\pi}{x^2 + \Gamma_i(\omega)} \right] d\omega' \right]. \qquad (4)$$

The summation is over the excited thresholds $E_i$ in the $N$-level wavefunction expansion. We associate the photon energy with the effective quantum number relative to each threshold $\omega' \to \nu_i$. Then the total width is calculated as follows:

$$\begin{aligned}\Gamma_i(\omega,\nu,T,N_e) &= \Gamma_c(i,\nu,\nu_c) + \Gamma_s(\nu_i,\nu_s^*) \\ &+ \Gamma_d(A,\omega) + \Gamma_f(f-f;\nu_i,\nu_i'),\end{aligned} \qquad (5)$$

where the partial broadening widths are due to collisional $\Gamma_c$, Stark $\Gamma_s$, Doppler $\Gamma_d$, and free–free transition $\Gamma_f$, respectively. The approximations outlined above should be valid, since collisional $\Gamma_(x)$ profile wings extend by $x^{-2}$ and are much wider compared to the shorter range of $exp(-x^2)$ for thermal Doppler and of $x^{-5/2}$ for Stark broadening. The following expressions are used for the computations:

$$\Gamma_c(i,\nu) = 5\left(\frac{\pi}{kT}\right)^{1/2} a_o^3 N_e G(T,z,\nu_i)(\nu_i^4/z^2), \tag{6}$$

where temperature, electron density, ion charge, and atomic weight are denoted as T, $N_e$, z, and A, respectively. The effective quantum number $\nu_i$ is calculated relative to the energy of each excited core ion $i : \omega \equiv E = E_i - \nu_i^2/z^2$ is a continuous variable. A temperature-dependent Gaunt factor

$$G(T,z,\nu_i) = \sqrt{3}/\pi[1/2 + ln(\nu_i kT/z)]. \tag{7}$$

is used as in [31] for treatment of the Stark effect.

The Stark width is related to the Mihalas–Hummer–Däppen equation-of-state and is $\approx (3F/z)n^2$, where F is the plasma electric microfield [11]. Generally, the dominant ion perturbers are protons with equal density to electrons, $N_e = N_p$; we take $F = [(4/3)\pi a_o^3 N_e)]^{2/3}$. Then

$$\Gamma_s(\nu_i, \nu_s^*) = [(4/3)\pi a_o^3 N_e]^{2/3} \nu_i^2. \tag{8}$$

In addition, a Stark ionization parameter $\nu_s^* = 1.2 \times 10^3 N_e^{-2/15} z^{3/5}$ is used when AI resonances are fully dissolved into the continuum for $\nu_i > \nu_s^*$ (as the Inglis–Teller series limit for line broadening [32]).

The Doppler width is

$$\Gamma_d(A,T,\omega) = 4.2858 \times 10^{-7} \sqrt{(T/A)}, \tag{9}$$

where $\omega$ is *not* the usual line center but taken to be each AI resonance energy. The free–free $\Gamma_f$ term in Equation (6) accounts for transitions among autoionizing levels with $\nu_i, \nu_i'$.

$$X_i + e(E_i, \nu_i) \longrightarrow X_i' + e'(E_i', \nu_i'). \tag{10}$$

The free–free transition probabilities for levels $E_i, E_i' > 0$ are also computed using RM codes [14].

The Rydberg series of AI resonances correspond to $(S_i L_i J_i)$ $n\ell$, $n \leq 10, \ell \leq 9$, with the effective quantum number defined as a continuous variable $\nu_i = z/\sqrt{(E_i - E)}$ $(E > 0)$, up to the highest core level, and are resolved for all cross-sections at ∼40,000–45,000 energies [15,17].

The otherwise unperturbed and unbroadened R-matrix cross-sections of an isolated atom become temperature- and density-dependent in HED plasmas. The new autoionization broadening method has been utilized to re-process all bound–free cross-sections at each temperature and density for all levels of each ion before opacity computations. The magnitude of this task is exemplified in Figure 2, which shows a sample of a single excited state of Fe XVIII at a range of densities along one isotherm corresponding to the temperature of the BCZ, $T = 2 \times 10^6$ K.

As described in [16], the main plasma broadening effects are thermal Doppler, ion Stark microfields, and electron collisions. As the plasma density increases, Stark and collisional broadening dominate, with the latter surpassing the former at very high densities, for example, at the BCZ density $N_e = 10^{23}$ cm$^{-3}$ in Figure 2 (see [16] for details).

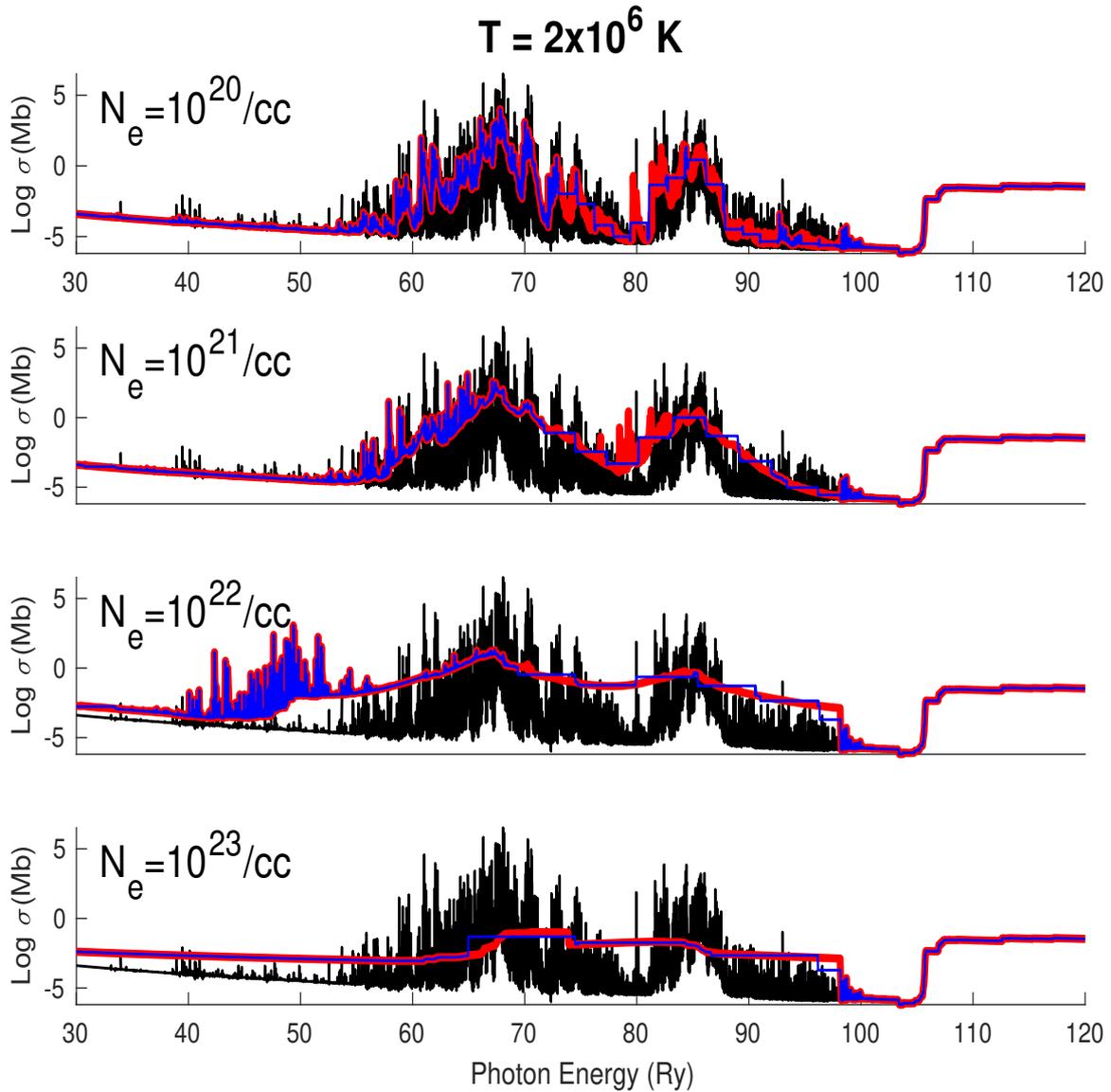

**Figure 2.** An example of RMOP plasma-broadened photoionization cross-sections for Fe XVIII of an excited bound level $2s^2 2p^4\ [^3P^e_0]\ 5s(^2P_{1/2})$ (ionization energy: 13.79 Ry) along the isotherm $T = 2 \times 10^6$ K and with electron densities of $N_e = 10^{20,21,22,23}$/cc: black—unbroadened RMOP cross-sections treated with plasma broadening effects; red—broadened; blue—broadened with Stark ionization cut-off [16]. Rydberg series of autoionizing resonance complexes broaden and shift with increasing density.

## 4. Monochromatic and Mean Opacities

Once the R-matrix data are computed for individual ions, and the plasma effects included, the monochromatic and mean opacities may be computed [10]. The monochromatic opacity comprises bound–bound (bb), bound–free (bf), free–free (ff), and photon scattering (sc) contributions:

$$\kappa_{ijk}(\nu) = \sum_k a_k \sum_j x_j \sum_{i,i'}[\kappa_{bb}(i,i';\nu) + \kappa_{bf}(i,\epsilon i';\nu) + \kappa_{ff}(\epsilon i,\epsilon' i';\nu) + \kappa_{sc}(\nu)], \qquad (11)$$

where $a_k$ is the abundance of element $k$, $x_j$ is the $j$ ionization fraction, $i$ and $i'$ are the initial bound and final bound/continuum states of the atomic species, and $\epsilon$ represents the electron energy in the continuum. Given the Planck function

$$B_\nu(T) = \frac{(2h\nu^3/c^2)}{e^{h\nu/kT} - 1}, \tag{12}$$

and given that the Planck Mean Opacity (PMO) is related to the total integrated photon flux absorbed,

$$\kappa_P B(T) = \int \kappa_\nu B_\nu d\nu. \tag{13}$$

The Rosseland Mean Opacity (RMO), $\kappa_R$ RMO, is related to the transmitted photon flux and is defined by the harmonic mean of the monochromatic opacity ($g(u) = dB_\nu/dT$, where $u \equiv h\nu/kT$).

$$\frac{1}{\kappa_R} = \frac{\int_0^\infty g(u)\kappa_\nu^{-1} du}{\int_0^\infty g(u) du} \quad ; \quad g(u) = u^4 e^{-u}(1 - e^{-u})^{-2}. \tag{14}$$

Close to the solar BCZ and the Z-experiment conditions, the iron opacity of about ∼85% is due to three ions, Fe XVII, Fe XVIII, and Fe XIX. Figure 3 shows the RMOP results for these ions under Sandia Z conditions and compared to the OP data. The differences between RMOP and OP are mainly due to autoionizing resonances and consequent redistribution of differential oscillator strengths.

The main effect in HED plasmas is to raise the background opacity due to dissolution of resonance structures with increasing electron density [16].

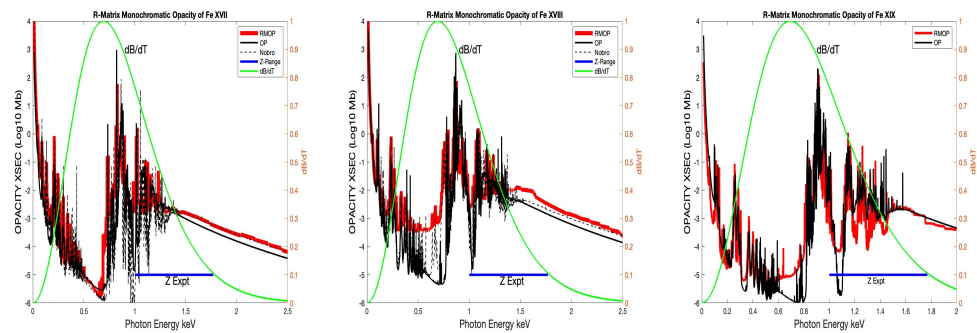

**Figure 3.** Monochromatic opacities of Fe XVII, Fe XVIII, and Fe XIX at $T = 2.1 \times 10^6$ K and an electron density of $N_e = 3.16 \times 10^{22}$/cc. Legends include RMOP results, including with (red) and without (dashed black) plasma broadening; OP results (black); Sandia Z energy range (blue); and dB/dT (green). Despite large enhancements owing to opacity redistribution by plasma effects between RMOP and OP, the integrated values agree to within 5%, confirming conservation of total oscillator strength.

## 5. Comparison with Experiments

One of the surprising results in the Sandia Z experiments is that the measured background opacity is much higher than all theoretical opacity model values in the energy region where there are sparse features due to the main contributing iron ions Fe XVII, Fe XVIII, and Fe XIX [19–22]. The RMOP results in Figure 3 have generally higher backgrounds and lower peaks than the OP results, filling in the opacity windows or dips, which are in better qualitative agreement with the Sandia Z measurements [18,19]. However, we did not find a background enhancement in iron opacity experimentally [18]. Theoretically, this is expected, since the RMOP and OP results should agree in continuum energy regions that are not dominated by resonances, since R-matrix and DW methods should yield the same background cross-sections if channel coupling is neglected [10], as shown in Figure 4. The comparison of the new but preliminary RMOP results compared with the Sandia experiment in Figure 4 shows that whereas RMOP opacities are in better agreement with the measured spectra overall than those of the OP, with enhanced backgrounds and shallower opacity "windows", this is not always the case, because the plasma broadening functional may redistribute opacity both towards higher- and lower-energy regimes [16]. Therefore, detailed calculations are necessary to determine whether the RMO would be enhanced at a given temperature–density combination or not; this would depend on where the peak of the Planck function lies relative to the redistributed opacity spectrum owing to plasma effects (a recent and comprehensive discussion of experimental results from opacity experiments on ICF plasma devices is given in [22].

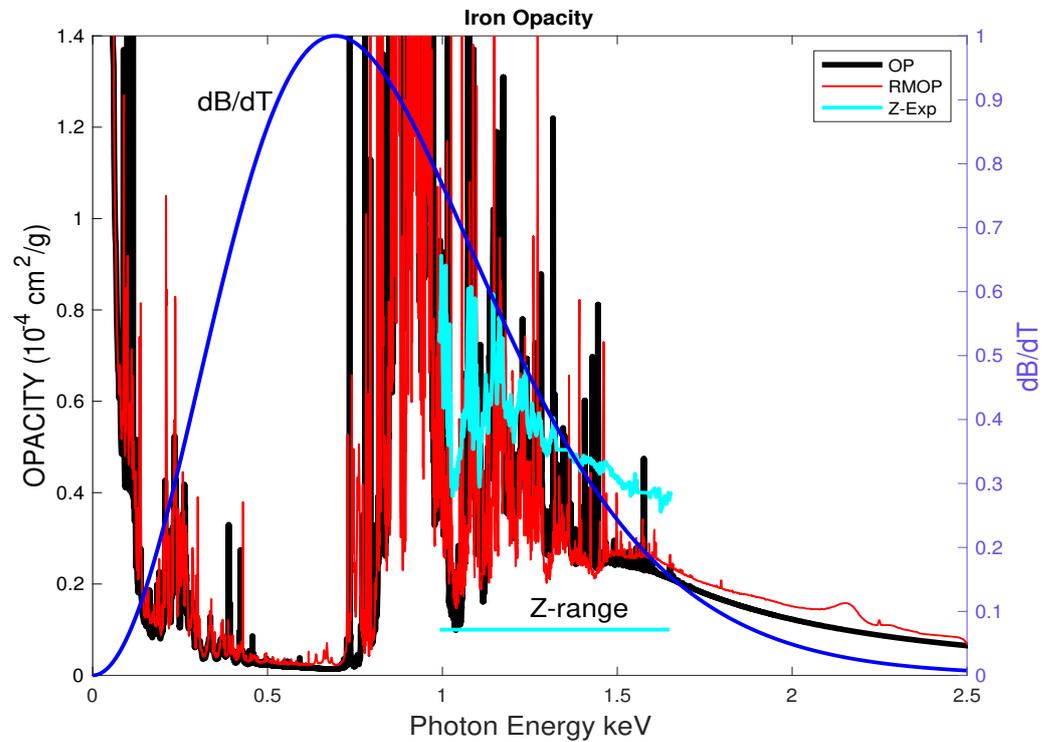

**Figure 4.** New iron monochromatic R-matrix opacity results (red) compared with the OP (black) and Sandia Z (cyan) experimental data. The detailed RMOP features show shallower opacity windows or dips and are in better qualitative agreement with the experiment, but the Z background disagrees with all theoretical models [19,20]. The Z measurements cover a relatively small energy range above the peak of the iron opacity spectrum, close to solar BCZ conditions.

However, an important point needs to be made. While the background continuum opacity may not be enhanced much, an increase in overall opacities cannot be ruled out. Indeed, based on the preliminary RMOP results shown herein, the RMOs could be much higher relative to those in the other opacity models, thereby ameliorating, if not resolving, the solar problem.

## 6. Equation-of-State

A possible source of uncertainty in opacity calculations is inaccuracy in the "chemical picture" MHD-EOS implemented in OP and RMOP calculations, with an improved variant labeled Q-MHD [8,11,18,27,33–36]. The MHD-EOS yields ionization fractions and level populations by introducing an occupation probability for each level owing to plasma microfields and consequent departure from the canonical Boltzmann–Saha equations in LTE.

The contribution to opacity from an atomic level population is

$$N_{ij} = \frac{N_j g_{ij} w_{ij} e^{-E_{ij}/kT}}{U_j}, \tag{15}$$

where $w_{ij}$ is the occupation probability of level $i$ in ionization state $j$, $E_{ij}$ is the excitation energy, and $U_{ij}$ is the internal partition function. The RMO, computed with two different versions of MHD-EOS, is highly sensitive to the distribution of level populations, but appears to "converge" after a certain number of levels for different ions. Figure 5 shows this apparent convergence of the three Fe ions of interest at solar BCZ temperatures and densities. Hundreds of levels are required for convergence, and they vary considerably across atomic species; this is expected, but indicates a "cut-off" beyond which RMOs remain constant with respect to the increasing number of levels. However, it also raises the question of whether this EOS cut-off and consequent redistribution of normalized level populations accurately reflects the interface between actual opacity and the extensive amount of atomic data, up to a thousand levels or more, included in opacity computations. (As explained in [18], the atomic data for levels are incorporated into MHD-EOS calculations according to the energy order within each spin-angular-parity symmetry $SLJ\pi$, and not absolute energy order. Nevertheless, the overall convergence trend should remain the same). Such a cut-off in MHD-EOS should not therefore be implemented without a more thorough examination and comparison among available EOS formulations to ascertain their precise limitations [36]. This is currently being carried out in another study.

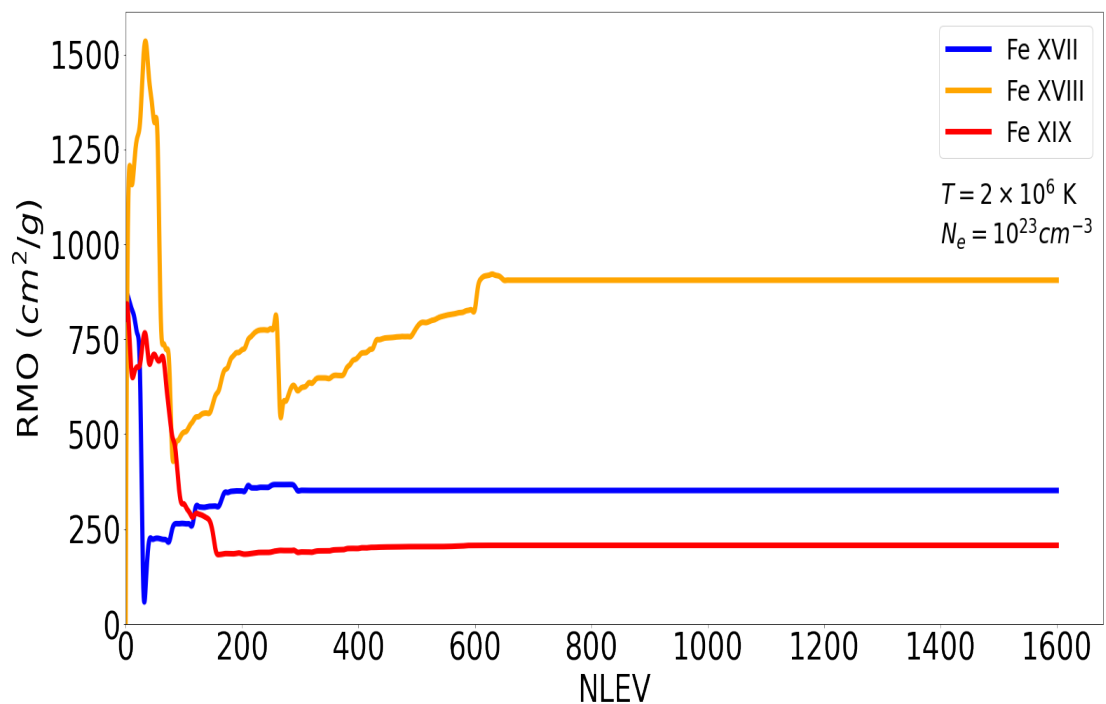

**Figure 5.** Rosseland Mean Opacity vs. number of levels in the equation-of-state included in RMOP computations for iron ions under solar BCZ conditions. Although the RMOs appear to 'converge' after a certain number of levels, there are huge variations, necessitating a closer re-examination of the QMHD-EOS interface with atomic data.

## 7. Conclusions

Definitive results for the opacities of astrophysically important heavy elements have not yet been obtained either theoretically or experimentally. We have described the continuation and extension of the Opacity Project based on the R-matrix method as originally envisaged. A few outstanding issues need to be resolved before sufficiently accurate opacities are obtained to solve the solar problem in particular, and HED plasma sources in general. These include inter-comparison of atomic data computed with various methods with respect to convergence, completeness, plasma effects in HED environments, the equation-of-state, and outstanding discrepancies with and among experimental data. Independently, helioseismic data could prove to be a useful factor for determining opacities and needs to be considered, as it pertains to various astrophysical parameters in the Sun [1].

From the point of view of improved atomic physics, the RMOP is an extension of the OP that aims to ensure that R-matrix calculations exhibit completeness, consistency, and convergence [17]. However, R-matrix calculations are highly intensive and require considerable effort compared to atomic-structure and DW calculations. Initial efforts have been focused on iron opacity calculations along the solar radius–temperature–density track, to be compared with experimental results on the one hand [19–21], and helioseismic data and results on the other hand [1]. More generally, the atom–plasma physical effects discussed herein are likely to manifest in HED plasma sources.

**Author Contributions:** Concptualization and computations, A.K.P. and S.N.N.; Opacities calculations, A.K.P.; Atomic calculations, S.N.N. All authors have read and agreed to the published version of the manuscript.

**Funding:** This work was partly supported by a grant from the U.S. National Science Foundation, AST-2407470.

**Data Availability Statement:** All data will be made available via the online database NORAD [28,29].

**Acknowledgments:** The computational work was carried out at the Ohio Supercomputer Center and the Ohio State University College of Arts and Sciences Computational Cluster. We would like to thank Divya Chari, Regner Trampedach, and Gaël Buldgen for engaging in discussions.

**Conflicts of Interest:** The authors declare no conflicts of interest.

# References

1. Buldgen, G.; Pain, J.; Cossé, P.; Blancard, C.; Gilleron, F.; Pradhan, A.K.; Fontes, C.J.; Colgan, J.; Noels, A.; Christensen-Dalsgaard, J.; Deal, M.; et al. Helioseismic inference of the solar radiative opacity. *Nat. Commun.* **2025**, *16*, 693. [CrossRef] [PubMed]
2. Asplund, M.; Amarsi, A.M.; Grevesse, N. The chemical make-up of the Sun: A 2020 vision. *Astron. Astrophys.* **2021**, *653*, A141. [CrossRef]
3. Basu, S.; Antia, H.N. Helioseismology and solar abundances. *Phys. Rep.* **2008**, *457*, 217–283. [CrossRef]
4. Christensen-Dalsgaard, J.; Di Mauro, M.P.; Houdek, G.; Pijpers, F. On the opacity change required to compensate for the revised solar composition. *Astron. Astrophys.* **2009**, *494*, 205–208. [CrossRef]
5. Seaton, M.J. Atomic data for opacity calculations. I. General description. *J. Phys. B* **1987**, *20*, 6363. [CrossRef]
6. Berrington, K.A.; Burke, P.G.; Butler, K.; Seaton, M.J.; Storey, P.J.; Taylor, K.T.; Yan, Y. Atomic data for opacity calculations. II. Computational methods. *J. Phys. B* **1987**, *20*, 6379. [CrossRef]
7. Seaton, M.J.; Yu, Y.; Mihalas, D.; Pradhan, A.K. Opacities for stellar envelopes. *Mon. Not. R. Astr. Soc.* **1994**, *266*, 805. [CrossRef]
8. The Opacity Project Team. *The Opacity Project*; IOP Publishing: Bristol, UK, 1995; Volume 1.
9. Burke, P.G. *R-Matrix Theory of Atomic Collisions*; Springer: Berlin/Heidelberg, Germany, 2011.
10. Pradhan, A.K.; Nahar S.N. *Atomic Astrophysics and Spectroscopy*; Cambridge Unversity Press: Singapore, 2011.
11. Mihalas, D.; Hummer, D.G.; Däppen, W. The equation of state for stellar envelopes. II-Algorithm and selected results. *Astrophys. J.* **1988**, *331*, 815–825. [CrossRef]
12. Nahar, S.N.; Pradhan, A.K. Large Enhancement in High-Energy Photoionization of Fe XVII and Missing Continuum Plasma Opacity. *Phys. Rev. Lett.* **2016**, *116*, 235003. [CrossRef]
13. Nahar, S.N.; Pradhan, A.K. Nahar and Pradhan Reply. *Phys. Rev. Lett.* **2016**, *116*, 235003. [CrossRef]
14. Pradhan, A.K.; Nahar, S.N.; Eissner, W. R-matrix calculations for opacities: I. Methodology and computations. *J. Phys. B* **2024**, *57*, 125001. [CrossRef]
15. Nahar, S.N.; Eissner, W.; Zhao, L.; Pradhan, A.K. R-Matrix calculations for opacities: II. Photoionization and oscillator strengths of iron ions Fe XVII, Fe XVIII and Fe XIX. *J. Phys. B* **2024**, *57*, 125002.
16. Pradhan, A.K. R-matrix calculations for opacities: III. Plasma broadening of autoionizing resonances. *J. Phys. B* **2024**, *57*, 125003. [CrossRef]
17. Zhao, L.; Nahar, S.N.; Pradhan, A.K. R-matrix calculations for opacities: IV. Convergence, completeness, and comparison of relativistic R-matrix and distorted wave calculations for Fe xvii and Fe xviii. *J. Phys. B* **2024**, *57*, 125004. [CrossRef]

18. Pradhan, A.K. Interface of equation of state, atomic data, and opacities in the solar problem. *Mon. Not. R. Astr. Soc.* **2024**, *527*, L179–L183. [CrossRef]
19. Bailey, J.; Nagayama, T.; Loisel, G.P.; Rochau, G.A.; Blancard, C.; Colgan, J.; Cosse, P.; Faussurier, G.; Fontes, C.J.; Gilleron, F.; et al. A higher-than-predicted measurement of iron opacity at solar interior temperatures. *Nature* **2015**, *517*, 56–59. [CrossRef]
20. Nagayama, T.; Bailey, J.E.; Loisel, G.P.; Dunham, G.S.; Rochau, G.A.; Blancard, C.; Colgan, J.; Cosse, P.; et al. Systematic Study of L-Shell Opacity at Stellar Interior Temperatures. *Phys. Rev. Lett.* **2019**, *122*, 235001. [CrossRef]
21. Hoarty, D.; Morton, J.; Rougier, J.C.; Rubery, M.; Opachich, Y.P; Swatton, D., Richardson, S.; Heeter, R.F.; McLean, K.; Rose, S.J.; Perry, T.S.; Remington, B. Radiation burnthrough measurements to infer opacity at conditions close to the solar radiative zone-convective zone boundary. *Phys. Plasmas* **2023**, *30*, 063302.
22. Heeter, R.F.; Hansen, S.B.; Johns, H.M.; Nagayama, T.; Aberg, D.P; Bailey, J.E.; Dutra, E.C.; Fontes, C.J.; Hohenberger, M.; Loisel, G.P.; et al. Technical Report: Milestone 7720: National Opacity Program—Tri-Lab Assessment of Measurements and Models. Available online: https://www.osti.gov/biblio/2430172 (accessed on 30 October 2023).
23. Pradhan, A.K. Plasma effects on resonant phenomena. *Can. J. Phys.* **2024**, *103*, 27–33. [CrossRef]
24. Berrington, K.A.; Eissner, W.; Norrington, P.N. RMATRX1: Belfast atomic R-matrix codes. *Comput. Phys. Commun.* **1995**, *92*, 290–420. [CrossRef]
25. Grant, I.; Quiney, H. GRASP: The Future? *Atoms* **2022**, *10*, 108. [CrossRef]
26. Mendoza, C.; Seaton, M.J.; Buerger, P.; Bellorin, A.; Melendez, M.; Gonzalez, J.; Rodriguez, L.S.; Palacios, E.; Pradhan, A.K.; Zeippen, C.J. OPserver: Interactive online computations of opacities and radiative accelerations. *Mon. Not. R. Astr. Soc.* **2007**, *378*, 1031–1035. [CrossRef]
27. Seaton, M.J.; Badnell, N.R. A comparison of Rosseland-mean opacities from OP and OPAL. *Mon. Not. R. Astr. Soc.* **2004**, *354*, 457–465. [CrossRef]
28. Nahar , S.N. Database NORAD-Atomic-Data for atomic processes in plasma. *Atoms* **2020**, *8*, 68. [CrossRef]
29. Nahar, S.N. Enhancement of the NORAD-Atomic-Data Database in Plasma. *Atoms* **2024**, *12*, 22. [CrossRef]
30. Delahaye, F.D.; Ballance, C.P.; Smyth, R.T.; Badnell, N.R. Quantitative comparison of opacities calculated using the R-matrix and distorted-wave methods: Fe XVII. *Mon. Not. R. Astr. Soc.* **2021**, *508*, 421–432. [CrossRef]
31. Dimitrijevic, M.S.; Konjevic, N. Simple estimates for Stark broadening of ion lines in stellar plasmas. *Astron. Astrophys.* **1987**, *172*, 345–349.
32. Inglis, D.R.; Teller, E. Ionic Depression of Series Limits in One-Electron Spectra. *Astrophys. J.* **1939**, *90*, 439. [CrossRef]
33. Badnell, N.R.; Seaton, M.J. On the importance of inner-shell transitions for opacity calculations. *J. Phys. B* **2003**, *36*, 4367. [CrossRef]
34. Badnell, N.R.; Bautista, M.A.; Butler, K.; Delahaye, F.; Mendoza, C.; Palmeri, P.; Zeippen, C.J.; Seaton, M.J. Updated opacities from the Opacity Project. *Mon. Not. R. Astr. Soc.* **2005**, *360*, 458–464. [CrossRef]
35. Nayfonov, A.; Däppen, W.; Hummer, D.G.; Mihalas, D. The MHD equation of state with post-Holtsmark microfield distributions. *Astrophys. J.* **1999**, *526*, 451. [CrossRef]
36. Trampedach, R.; Däppen, W.; Baturin, V.A. A synoptic comparison of the Mihalas-Hummer-Däppen and OPAL equations of state. *Astrophys. J.* **2006**, *646*, 560.